\documentclass[aps,prb,twocolumn,showpacs]{revtex4}

\usepackage{graphicx}
\usepackage{dcolumn}
\usepackage{bm}
\usepackage{amssymb}

\usepackage{xcolor}

\begin{document}

\title{Graphene single electron transistor  as a spin sensor for  magnetic adsorbates}

\author{J. W. Gonz\'alez$^{(1)}$,F. Delgado$^{(1)}$, and J. Fern\'andez-Rossier$^{(1,2)}$}
\affiliation{$^{(1)}$ International Iberian Nanotechnology Laboratory (INL),
Av. Mestre Jos\'e Veiga, 4715-330 Braga, Portugal
\\ $^{(2)}$ Departamento de F\'isica Aplicada, Universidad de Alicante, 03690 San Vicente del Raspeig, Spain}

\date{\today}

\begin{abstract}

We study single electron transport through a graphene quantum dot with magnetic adsorbates. 
We focus on the relation between the spin order of the adsorbates and the linear conductance of the device.
The electronic structure of the graphene dot with magnetic adsorbates is modeled  through numerical diagonalization of a tight-binding model with an exchange potential. 
We consider  several mechanisms by  which the adsorbate magnetic state can influence 
transport in a single electron transistor: by tuning the addition energy,  
by changing the tunneling rate and, in the case of spin polarized electrodes, through magnetoresistive effects.  
Whereas the first  mechanism is always present, the others require that the electrode has either 
an energy or spin dependent density of states. We  find that graphene dots are optimal systems to detect the spin state of a few magnetic centers. 

% 73.23.Hk Single-electron tunneling
% graphene, 81.05.ue
% magnetic field sensors, 85.75.Ss
% transport graphene, 72.80.Vp
\end{abstract}

\maketitle

\section{Introduction}

Graphene is a very promising candidate for high precision molecular sensing, 
due to  its  extremely large surface to volume ratio, and its electrically 
tunable large conductivity.\cite{Schedin07,Wernsdorfer_NL,Pisana}%,Patent} 
On the other hand, being a zero-gap semiconductor with small mass and small 
density of spinfull nuclei, makes graphene a  material with potentially 
large spin lifetime both, for carriers and host magnetic dopants. \cite{Pesin2012}
Taken together, these two ideas naturally lead to the use of graphene as a 
detector of the spin state of extrinsic magnetic centers, in the form of magnetic 
adatoms, vacancies and spinfull molecules.  
This connects with recently reported \cite{Wernsdorfer_NL,Candini-PRB-2011} 
experiments in which gated graphene nanoconstrictions, operating in the 
single electron transport (SET) regime, showed hysteresis in the linear 
conductance when a magnetic field is ramped. This behavior was observed 
both when the molecular magnets were intentionally deposited 
on graphene,  as well as in carbon nanotubes,\cite{Wernsdorfer_NatMat}
but also in the case of  bare graphene nanojunctions,\cite{Candini-PRB-2011}
where some type  of graphene local moments \cite{Yazyev07,Palacios2008,Soriano2010} 
is probably playing a role.   
  
\begin{figure}[ht!]
\includegraphics[clip,width=0.49\textwidth,angle=0]{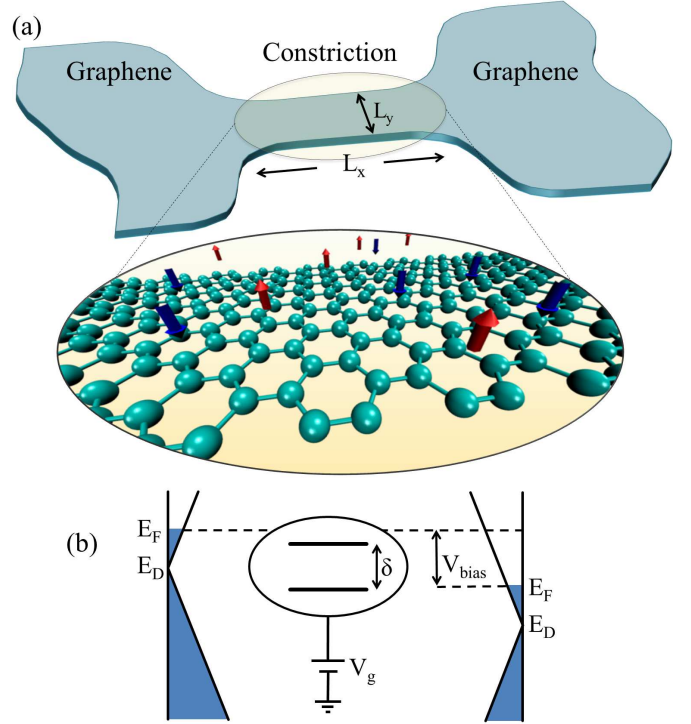} %width=6.5cm
\caption{(Color online) (a) Scheme of a graphene
constriction with randomly distributed
magnetic centers. (b) Diagram with the
system energy levels and graphene
density of states.
}
\label{Fig:Esq} \end{figure}

The graphene spin sensor experiments of Ref.~\onlinecite{Wernsdorfer_NL}, % \onlinecite{Candini-PRB-2011}
are performed in the  Coulomb Blockade regime, showing a vanishing 
linear conductance except in the neighborhood of specific values of the gate 
voltage $V_g$.  This means that the graphene nanoconstriction is weakly 
coupled to the electrodes and has a charging energy 
larger than the thermal energy ($T \sim 100$ mK). 
The  height of the linear conductance peaks is significantly smaller than 
$G_0=2e^2/h$, the quantum of conductance.
These conditions imply that transport 
takes place in the sequential regime.\cite{Beenakker}
Thus,  current flow takes place due to sequential tunneling  
of electrons through the graphene constriction, which we refer to as 
the central region in the rest of the paper, and the entire device behaves 
like a single electron transistor.\cite{Geim2008,review-dot}

The aim of this work is to  provide a theoretical background to understand 
how the magnetic state of localized magnetic moments affects transport through the 
graphene nanoconstriction in the SET regime. This is different from previous works 
where the influence of the magnetic state of magnetic edges \cite{Fede2009} and 
adsorbed hydrogens \cite{Soriano2010} on the conductivity was studied in the ballistic 
regime, with a central island strongly coupled to the electrodes, and also in 
the diffusive regime,\cite{Castro-Neto} as well as SET through graphene islands with magnetic zigzag edges.\cite{Ezawa}

The paper is organized as follows. 
In Sec. \ref{secmodel} we discuss a tight-binding Hamiltonian for the  graphene island exchanged coupled 
to the spins of the
 magnetic adsorbates.
The results of this microscopic calculation justify the use of a simple single-orbital spin-split model for the SET, discussed in
section  \ref{sec2}, together with the  possible mechanisms that enable the magnetic sensing. 
Finally, conclusions are presented in Sec.~\ref{conclu}.

\section{Model for graphene island with magnetic adsorbates\label{secmodel}}

\subsection{Hamiltonian}
Our starting point is a microscopic model for  electrons confined in a graphene nanoisland which are exchanged coupled to the magnetic centers.
The graphene central island  is described with  a 
tight-binding Hamiltonian for the honeycomb lattice contained in a rectangular 
stripe of dimensions $L_x\times L_y$. In order  to avoid the spin-polarized  states formed at the the zigzag edges,\cite{Yazyev,JP_JR} we impose periodic boundary conditions in one direction so that the structure only has  open edges of  armchair type.     
The coupling to the magnetic moments of the adsorbate molecules $\vec{m}(i)$ is then assumed to be 
a local exchange $J$ or spin-dependent potential,  affecting $N$ sites
randomly selected in the graphene central island.  
For simplicity, we consider that the magnetic moments of the molecules are all oriented along 
the same axis, which we choose as the spin quantization axis.  
These assumptions are good approximations in the case of strongly uniaxial TbPC$_2$ molecules.\cite{Wernsdorfer_NL}  Hence, we can write the Hamiltonian of the graphene and adsorbates as
\begin{equation}
\mathcal{H}  = \mathcal{H}_0 + J \sum_i m_z(i) S_z(i) + eV_g \left(N_{TOT}-\hat{N}\right),
  \label{hamil}
\end{equation}
%H_0 =t \sum \left( a^{\dagger}_{\sigma,i}  b_{\sigma,j} + h.c. \right)
where $\mathcal{H}_0$ is the tight-binding Hamiltonian for $\pi$-electrons in 
graphene considering nearest neighbor interactions, 
$J$ is  the strength of the exchange 
coupling between the graphene electrons and the magnetic moment of the molecules,
which can take two values, 
$m_z(i)=\pm 1$.   Finally, the last term in the Hamiltonian  describes the electrostatic coupling of the total charge of the dot, which can be either $0$ or $e$,  given by the difference in the number of electrons $\hat{N}$ and the number of carbon atoms $N_{\rm TOT}$ in the central island.
$S_z(i)$ is the local spin density of the $p_z$ electrons in graphene at site $i$,
\begin{equation}
S_z(i)=
\frac{1}{2} \left( 
c^{\dagger}_{i\uparrow}c_{i\uparrow}
-c^{\dagger}_{i\downarrow}c_{i\downarrow} 
\right),
\end{equation}
where  $c^{\dagger}_{i\uparrow}$ creates an $\uparrow$ electron at the $p_z$  orbital of site $i$ of graphene.
 In the following we assume that  magnetic fields controlling the spin orientation of the adsorbates, are applied along the plane of graphene so that it is a good approximation to neglect the diamagnetic coupling to the graphene electrons. 

%
%
%
%
%\subsection{Microscopic mechanisms}
There are several independent  microscopic  mechanisms for  spin dependent interaction between magnetic adsorbates
and the graphene $\pi$ electrons that can be modeled with Eq. (\ref{hamil}).
In the case of magnetic molecule such as TbPC$_2$,  used in in Ref. \onlinecite{Wernsdorfer_NL}, 
the magnetic Tb atom is separated from the graphene electrons by the non-magnetic atoms of the molecule, 
 and the most likely mechanism for spin coupling is  kinetic exchange.\cite{Anderson61}
This coupling will generate a local Kondo-exchange between graphene 
electrons and the molecules.\cite{Schrieffer-Wolff}
More complicated scenarios, like coupling of graphene electrons 
to unpaired electrons in the organic rings of the molecules, would 
imply that every molecule affects several sites in graphene.  
 Direct dipolar 
coupling would also affect several sites per molecule, but the average 
magnetic field created by a magnetic moment of 5 $\mu_B$ at 0.5 nm 
on a disk with an area  around 400 nm$^2$, the graphene constriction area in Ref.~\onlinecite{Wernsdorfer_NL},
is  smaller than 1 $\mu$T, which would yield a negligible maximal Zeeman coupling of neV per molecule.

%{\it  Microscopic model}

\subsection{Relevant energy scales}

The reported dimensions of the central region, $L_x\simeq L_y\simeq 20$ nm, 
lead to an energy spacing $\Delta \epsilon$  of the single particle spectrum 
much larger than the temperature and the charging energy.\cite{Geim2008,review-dot} We also assume that the exchange
 induced shifts are smaller than the single particle splitting. As a result, the effect of exchange is to shift the bare energy levels, without mixing them.
   Thereby, we can safely assume  that  electrons tunnel through  just one of the single particle levels, which might be spin-split  due to exchange with molecules.  
         We assume that the charge of the central island fluctuates between 
$q=0$ and $q=-|e|$,  and that the transport level is the  lowest unoccupied level of the central island spectrum. The energy of the transport level reads: 
\begin{equation}
\epsilon_T^{\sigma}= \epsilon_0 + \sigma\frac{\delta}{2} - |e| V_g,
\end{equation}
where $\epsilon_0$ is the single particle electron level, $\sigma = \pm 1$ denotes the spin direction  and
$\delta$ is the  magnitude of the spin splitting, which is a functional of the magnetic landscape $\{m_z\}$. 

\begin{figure}[t]
\includegraphics[clip,width=0.49\textwidth,angle=0]{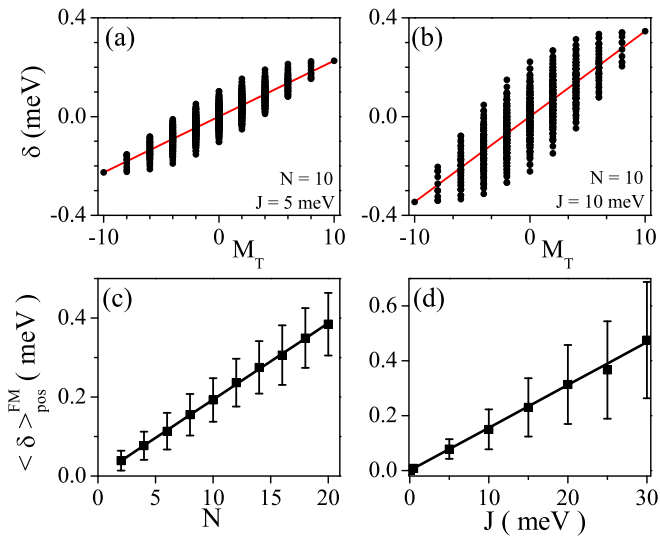} %width=6.5cm
\caption{Spin splitting $\delta$ for a sample of dimensions  $L_y = N_y a$ and 
$L_x = N_x \sqrt{3} a$, with $N_x=15$ and $N_y=17$, being $a=2.46 \AA$ 
the lattice constant of graphene. 
$\delta$ versus total magnetization $M_T$ for a single spatial distribution of $N=10$  magnetic impurities  
with $J=5$ meV (a) and $J=10$  (b). The red solid line indicates the average 
value $\langle \delta\rangle_{\rm pos}$. 
Average $\langle \delta\rangle_{\rm pos}^{\rm FM}$ versus number of magnetic molecules $N$ 
for $J=5$ meV (c) and exchange energy $J$ for $N=4$ (d).
The error bars correspond to the standard deviation.}
\label{Fig:Param} 
\end{figure}

Within our model, a given magnetic landscape is defined by the location $i$ and the magnetic state $m_z(i)$ of the $N$ magnetic adsorbates.  In Figs.  \ref{Fig:Param} (a, b), we plot the value of $\delta$ for all the possible magnetic states of a given arrangement of $N=10$ atoms, for two different values of $J$. This choice  corresponds to the estimated number of molecules in Ref. \onlinecite{Wernsdorfer_NL}.  These figures show a correlation between the magnitude of the splitting $\delta$ and the total magnetization $M_T$.  The dispersion of $\delta$ for a fixed total magnetization $M_T$ is the outcome of indirect exchange coupling\cite{Brey-Das-Sarma}  

For comparison with the experiments, it is worth considering two extreme magnetic landscapes. At large external field, all the 
magnetic moments are aligned, i.e.,  $m_z(i)= +1$. We refer to this as the 
ferromagnetic (FM) landscape.  At magnetic fields smaller than the 
coercive field of the magnetic molecules, their average magnetization 
should be zero and thus, $\sum_i m_z(i)=0$. We refer to these cases as  non-magnetic (NM).
In order to sample the positional disorder   we perform an average over positional  configurations, both for NM and FM cases.   For a fixed spin choice $\{m_z\}$ with $M_T=0$, an  average over positional configurations  yields $\langle \delta\rangle_{\rm pos}=0$. The reciprocal statement is also true: for a fixed positional configuration, an  average over all the magnetic landscapes with $M_T=0$ also yields an average $ \langle\delta\rangle_{\rm spins}=0$.

In Fig. \ref{Fig:Param}  we plot the average 
 $\langle\delta \rangle_{\rm pos}^{\rm FM}-\langle\delta \rangle_{\rm pos}^{\rm NM}=\langle\delta \rangle_{\rm pos}^{\rm FM}$ over $500$  
realizations  as a function of 
the number of molecules $N$ [Fig.~\ref{Fig:Param}(c)] and as a function of the molecule electron exchange $J$ [Fig.~\ref{Fig:Param}(d)].
We have also calculated $\langle\delta\rangle_{\rm pos}^{\rm FM}$  fixing the number of magnetic centers $N$, the strength of the coupling $J$ and changing $N_{\rm TOT}= N_x N_y$, the total number of carbon atoms in the island. 
We find that the results of all these simulations can be summarized in the following  equation:
\begin{equation}
\left\langle \delta\right\rangle_{\rm pos}^{\rm FM} \approx  \frac{N }{N_{\rm TOT}} J.
\label{du}
\end{equation}
Whereas this result has been obtained from exact numerical diagonalization of the Hamiltonian, this dependence can be rationalized  using first order perturbation theory, which yields the spin dependent shift  of the transport level:
 \begin{equation}
 \Delta\epsilon_T^{\sigma} =\frac{\sigma}{2} J \sum_i^{N} |\phi_T(i)|^2 m_i,
 \end{equation}
 where $\phi_T(i)$ is the $J=0$ wave function of the transport orbital.  We now use  $|\phi_T(i)|^2\simeq \frac{1}{N_{\rm TOT}} $ so that  we can approximate: 
 \begin{equation}
 \Delta\epsilon_T^{\sigma}\simeq \frac{\sigma}{2} \frac{J}{N_{\rm TOT}} M_T %\sum_i  m_i
 \end{equation}
Using 
the fact that $M_T=N$ for the FM  configurations and $0$   for the NM ones,   we arrive to Eq. (\ref{du}).

\section{Spin-Split Single Orbital Model for SET \label{sec2}}
In this section we discuss SET across a central island with a single spin split particle level. This is justified by the results of the previous section.  We obtain expressions for the current of the system and we    discuss the
conditions  under which the conductance depends on the magnetic state of the single electron transport. 

\subsection{Single electron transistor with a spin-split single orbital model}

 We consider single electron transport though a spin split single  transport level,\cite{Recher2000} with energy $\epsilon_T^\sigma$. 
 We assume that the occupation of the transport level can be either 0 or 1, the doubly occupied configuration being much higher in energy. 
Within these approximations, the  transport level  has  three relevant many body states: uncharged, 
and the two charged with $\uparrow$ or $\downarrow$ spins. 
In the zero-applied bias limit, each of these states will  
be occupied according to the thermal equilibrium distribution, which we
denoted as $P_0$, $P_{\uparrow}$ and $P_{\downarrow}$ respectively.  
In the SET regime we are interested and within the linear response 
($eV_{bias}\ll k_B T$), transport will be enabled only when the 
addition energy lies within the thermally broadened 
transport window defined by the applied bias.

Under these approximations, the current flowing from the left electrode to 
the central island is given by:  
\begin{equation}
I=e\sum_\sigma \left\{P_0 W_{0\to \sigma}^L-P_\sigma W_{\sigma\to 0}^L\right\},
\label{Eq:I}
\end{equation}
where $W_{0\to \sigma}^L$ and $W_{\sigma\to 0}^L$ are 
rates for electron tunneling  from the left electrode to the dot and vice-versa.  
Continuity equation ensures that this current is identical to the current flowing 
towards the right electrode and, thereby, equal to the net current flow.   
The tunneling rates for electron tunneling out of and into the dot  
\cite{Haug_Jauho_book_1996}  are given respectively by
\begin{equation}
W_{\sigma\to 0}^L=\frac{2\pi}{\hbar} |T_L|^2
 \rho_L(\epsilon_T^\sigma)
\left[1-f(\Delta_\sigma+V_{bias})\right],
\label{rate1}
\end{equation}
and
\begin{equation}
W_{0\to \sigma}^L=\frac{2\pi}{\hbar} |T_L|^2
 \rho_L(\epsilon_T^\sigma)
f(\Delta_\sigma+V_{bias}),
\label{rate2}
\end{equation}
where $T_L$ is the strength of the dot - left electrode coupling,  and
\begin{equation}
\Delta_{\sigma}\equiv \epsilon_T^\sigma -E_F,
%\epsilon_0 +\frac{ \sigma}{2}\delta  - eV_g- E_F
\end{equation}
are the spin dependent {\em addition energies}.   Importantly, both $\delta$ and $V_g$ appear on equal footing, as additive quantities
in this equation.
The density of 
states of the left electrode, evaluated at the spin-dependent transport level energy,  is denoted by  $\rho_L(\epsilon_T^\sigma)$, 
while $f(\epsilon_\sigma) = \left(e^{\beta \Delta_\sigma } +1 \right)^{-1}$ denotes the Fermi function.
The electrode Fermi energy $E_F$ is taken to change linearly with the bias voltage $V_{bias}$.  
In the zero bias limit, the linear conductance reads:
\begin{equation}
G= G_0 \sum_{\sigma} \left(P_0 + P_{\sigma}\right)
\frac{\Gamma_{\sigma}}{k_B T} {\rm Sech}^2\left(\frac{\beta\Delta_\sigma}{2}\right),
\label{Eq:G}
\end{equation}
where 
\begin{equation}
\Gamma_\sigma/\hbar=\frac{2 \pi}{\hbar} |T_L|^2 \rho_L(\epsilon_T^\sigma),
\label{lifeT}
\end{equation}
is the single particle tunneling rate between the electrode and the transport level.

\subsection{Influence of magnetic state on conductance}
\begin{figure}[t]
\includegraphics[clip,width=0.49\textwidth,angle=0]{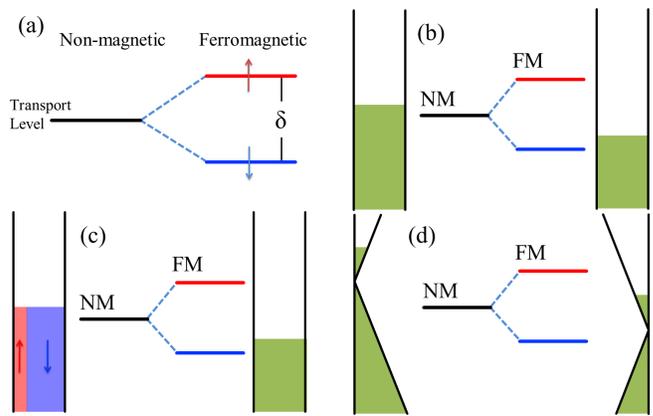} %width=6.5cm
\caption{(Color online)
(a) Scheme showing the transport level energy splitting.    Scheme of spin-dependence of transport due
to: (b) detuning  of the transport level with respect to the electrode Fermi energy, (c) magnetoresistance associated to
spin polarized electrode(s), and (d) energy dependent tunneling rates. }
\label{Fig:schems} \end{figure}

\begin{figure}[ht!]
\includegraphics[clip,width=0.49\textwidth,angle=0]{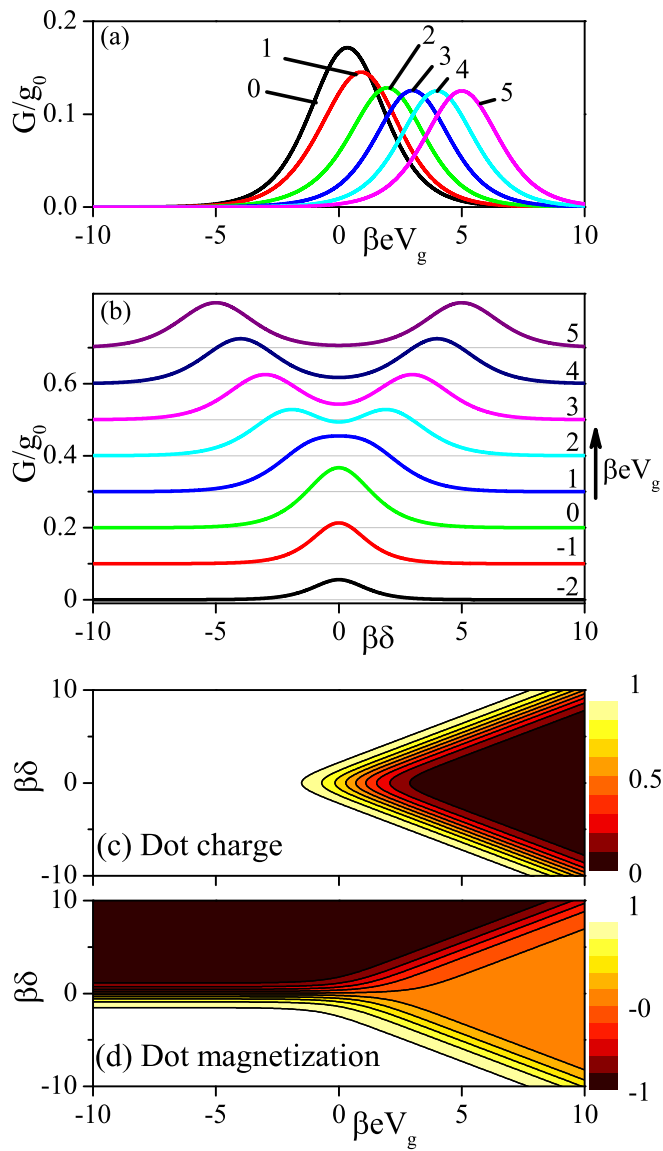} %width=6.5cm
\caption{(Color online) Conductance in units of $g_0=G_0\beta\Gamma$ as a function of the gate voltage $eV_g$ and the 
level splitting $\delta$. 
Panel (a) shows the normalized conductance as a function of the gate potential $e \beta V_g$,  where
the labels corresponds to the different $\beta \delta$ values.
Panel (b) shows  the conductance as a function of $\beta \delta$ for several values of the gate 
$e \beta V_g$. For the sake of clarity all curves have been displaced by $0.1$. 
In (c, d) we present a contour plot of the dot charge (defined as $Q = P_\uparrow + P_\downarrow$) 
and the dot magnetization ($m = P_\uparrow - P_\downarrow$) as a function of the gate voltage $e V_g$
and level splitting $\delta$.} 
\label{Fig:2} 
\end{figure}

From the above discussion %equations (\ref{add}) and (\ref{lifeT})
it is apparent that, for a given gate potential $V_g$ 
and temperature $T$, the linear conductance depends on the magnetic 
landscape affecting the central island through two classes of independent mechanisms, illustrated in Fig.~\ref{Fig:schems}: 
\begin{enumerate}

\item The change of the addition energies $\Delta_{\sigma}$  which, as we show below,  would result in a 
 lateral shift of the $G(V_g)$ resonance curve 
 [Fig.~\ref{Fig:2}(a)]. 
  
\item The change of the electron lifetime $\Gamma=\sum_{\sigma} \Gamma_{\sigma}$, 
 that would result in a vertical resizing of the $G(V_g)$ resonance curve 
 [Fig.~\ref{Fig:G_J_delta}(a)].
 
\end{enumerate}

 In the first mechanism the change in the magnetic state modifies the value of $\delta$, which must have a similar effect than changing the gate potential. It resembles the magneto-Coulomb effect,\cite{Ono,Vanwees} 
by which the applied magnetic field changes the Fermi energy  of the  
electrode, shifting the $G(V_g)$ curves.
However, this first mechanism necessarily implies a change of sign of the 
variation of $G$ as the gate potential is scanned along the resonance 
(see top panel of Fig. \ref{Fig:2}). 
Importantly, this is {\em not observed in the experiments} 
with magnetic molecules,\cite{Wernsdorfer_NL} but it is observed
in the case of graphene nanoconstrictions.\cite{Candini-PRB-2011}

Motivated by the behavior reported in Ref.~\onlinecite{Wernsdorfer_NL}, we pay attention also to the second mechanism.  For spin un-polarized transport, the change in the transport energy level results in a change
on tunneling rate $\Gamma/\hbar$ only if the electrode density of states depends on energy, which is exactly the case of graphene.  For spin-polarized transport, the relative orientation of the electrode and island magnetic moment give rise to magneto-resistive effects that are accounted for by the changes in $ \Gamma_{\sigma}$

%

%\section{Conductance for spin-split SET model }
\label{sec-results}

\subsection{Transport for constant tunneling rates\label{results1}}
We now  discuss our transport simulations  for  the graphene single electron transistor spin sensor.
We focus  on the  first spin sensing mechanism in a single electron transistor: changes
in spin splitting of the transport level produce changes in addition energies $\Delta_{\sigma}$ (figure  (\ref{Fig:schems})b).  For that matter,
we   neglect both the energy  and spin dependence of the tunneling rates $\Gamma_{\sigma}$.
In Fig. \ref{Fig:2}(a)  we show the linear conductance, in units of $g_0=\beta \Gamma G_0$,  as a function of gate voltage, for several values of the transport level splitting $\delta$, in units of $k_BT$. It is apparent that the  Coulomb Blockade peaks undergoes a lateral shift, as expected from the fact that $V_g$ and $\delta$ appear on equal footing on the spin dependent addition energies.   At $\Delta=0$ the two spin channels contribute. Therefore, as we increase $\delta$, the height of the conductance peaks decreases, because one of the two spin channels is removed from the transport window of width $k_BT$

 In figure (\ref{Fig:2})b we plot the variations in the linear conductances as a function of the spin splitting $\delta$, for several values of $V_g$. We see two types of curves.  For values of $V_g$  such that the transport level  is occupied, as we increase $\delta$ the transport level is pushed downwards, away from the elastic transport window,  switching off the transistor conductance.   In contrast, for $V_g$ such that the transport level is empty for $\delta=0$, lying above the elastic transport window,  ramping $\delta$ makes one of the two spin states of the transport level enter the transport window, giving rise to the double peak structure. 
The fact that $\delta$ and $V_g$ play analogous roles is illustrated in Figs.~\ref{Fig:2}(c, d), where we show the average magnetization and occupation of the transport level in the phase diagram defined by these two variables.

\subsection{Transport with energy  dependent tunneling rates\label{results2}}
We now consider the second mechanism for spin sensing  in a single electron transistor: changes
in spin splitting of the transport level produce changes in the tunneling rates $\Gamma_{\sigma}$. This can happen for two reasons: 
\begin{enumerate}
\item One of the  electrodes is spin polarized, so that $\Gamma_{\downarrow}\neq \Gamma_{\uparrow}$. Spin polarized transport is sensitive to the product of the magnetic moment of electrode and central island. This type of effect has been thoroughly discussed in the case of SET with ferromagnetic electrodes.\cite{Barnas98,Seneor2007}

\item The density of states of the electrode depends on  energy. Thus, changes in the transport level change  $\Gamma_{\sigma}$, for both spins.   This is a natural scenario for graphene electrodes. \cite{dots_Ensslin,Sols-Guinea} 
\end{enumerate}

\begin{figure}[t]
\includegraphics[clip,width=0.5\textwidth,angle=0]{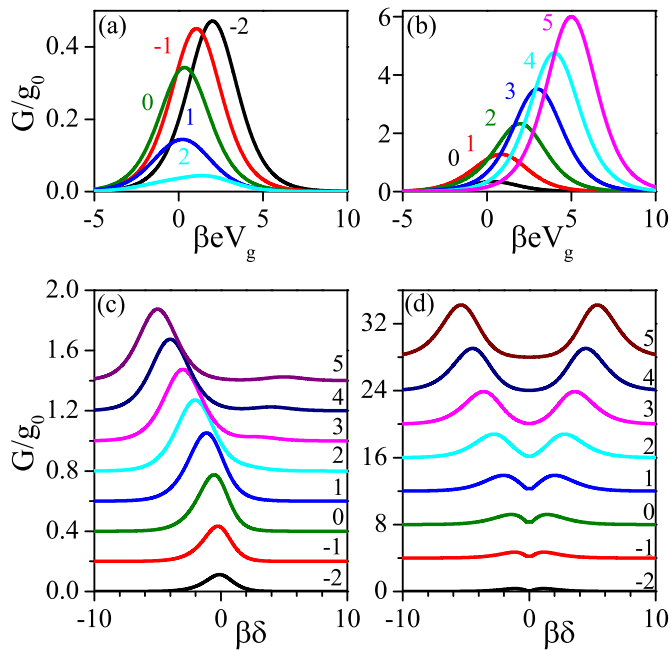} %width=6.5cm
\caption{(Color online) Normalized Conductance for ferromagnetic electrodes as a function of 
the level splitting $\delta$ for several gate values. 
In (a, c) the density of states depends of the  polarization $\rho = \rho_0 (1 + \sigma {\cal P})$, 
in this particular case we take a polarization of ${\cal P} = 0.9$.
In (b, d) the electrode density of states is linear with the energy, 
$\rho(\Delta_\sigma) = \rho (\epsilon_0+\sigma \delta/2-E_D)$.
For the sake of clarity the conductance curves in (c, d) have been displaced by $0.2$ 
and $4$ units respectively.} 
\label{Fig:G_J_delta} 
\end{figure}

Let us consider first  the  case of  spin polarized  electrodes.
We do the assumption that the density of states are spin dependent but energy independent:  $\rho_\sigma = \rho_0 (1 + \sigma {\cal P})$, with ${\cal P}$ the electrode polarization.  In Fig. \ref{Fig:G_J_delta}(a) we plot the linear conductance vs $V_g$ curves for several values of the splitting $\delta$, assuming a large value of the electrode spin polarization ${\cal P}=0.9$.  It is apparent that, on top of the shift of the resonance curve whose origin was discussed in the previous subsection, there is a change in the amplitude of the curve. Notice that $G(\delta=k_B T)$ and  $G(\delta=2 k_B T)$ are smaller than $G(0)$ for all values of $V_g$. In this specific sense, the gate-independent spin contrast is similar to the experimental report with magnetic molecules.\cite{Wernsdorfer_NL} 
 In Fig.~\ref{Fig:G_J_delta}(a)  we show the linear conductance as a function of  $\delta$ for different values of $V_g$.  It is apparent that, as opposed to the case of non-magnetic electrodes shown in Fig. 5(b),  the  function $G(\delta$) is no longer an even functions, reflecting the magneto-resistive behaviour. Basically, transport is favored when the spin polarization of the electrode and the central island are parallel.

We now consider a non-spin polarized electrode with an energy dependent density of states. 
This scenario  occurs naturally in graphene. If we consider idealized graphene electrodes, neglecting effects of interactions, disorder and confinement, we have  $\rho(\epsilon)=\rho_0|\epsilon-E_D|$, where $E_D$ is the Dirac point.
The $G(V_g)$ curves, shown in Fig. \ref{Fig:G_J_delta}(b) for different values of $\delta$,
 shift and change amplitude. The shift  is related to the change of the addition energies, discussed in the previous subsection, and the change in amplitude comes from the variation of the tunneling rate as the transport level scans the energy dependent density of states of the electrode.  
 
 In Fig. \ref{Fig:G_J_delta}(d) we plot $G(\delta)$ for several values of $V_g$.  The curves are similar to the case with energy independent tunneling rates, except for the dip at zero $\delta$ which occurs because we chose the bare transport level right at  Dirac point.  This is the most favorable choice to maximize the effect of energy dependence of $\Gamma$. From our results, and given the fact that experimentally is not possible to put the Fermi energy  arbitrarily close to the Diract point,\cite{Geim-closest}   we find it unlikely that this effect is playing a role in the experiments.

\subsection{Sensitivity  of the single electron transistor spin sensor}
 We now discuss the sensitivity of the spin sensor based on the graphene single electron transistor, as described by our model, neglecting changes in $\Gamma$.    From Fig.~\ref{Fig:2}(a) we propose, as rule of thumb, that variations of $\delta$ similar or larger than $k_BT$ can be resolved.  Estimating $\delta$ from the case of fully spin polarized magnetic adsorbates,  given in Eq. (\ref{du}), we find a relation between
 the minimal number of magnetic centers  $N$ % per unit area 
that can be detected, 
and the temperature and exchange constant: 
 \begin{equation}
\frac{N }{N_{\rm TOT}} > \frac{ k_B T}{J}.
 \label{sensitivity}
\end{equation}

It is apparent that decreasing the temperature, or increasing the spin-graphene exchange coupling, increases the  sensitivity of the device (makes it possible to detect a smaller concentration of molecules). For instance, at $100$ mK, and taking
  $N_{\rm TOT}\sim 15000$, which corresponds to an approximate area of $400$ nm$^2$,$\;$ one could detect $10$ molecules for an exchange coupling $J\gtrsim 15$  meV. 

Recent reports have shown that it is possible to fabricate graphene nano islands with lateral dimensions of 1nm.\cite{Barreiro2012}   These dots have $N_{\rm TOT}<100$.  Thus, they would permit  the detection of the spin of a  single magnetic adsorbate  provided that $k_BT$ is kept hundred times smaller than $J$.  For $T= 100$ mK this implies  $J>1$ meV.
Interestingly, in such a small dot  Coulomb Blockade persists even at room temperature,  but increasing  $k_BT$ keeping the sensitivity would require also to increase $J$.

%%%%%%%%%%%%%%%%%
\section{ Discussion and conclusions \label{conclu}}
%
%

% Classical spins
% Lack of back-action of transport
% No spin relaxation of quasiparticles
% No correlations
%No valley
 
Because of its  structural and electronic properties,  graphene is optimal for a spin sensor device.  Being all surface, 
the influence of  adsorbates on transport should be larger than any other bulk material. Because of the linear relation momentum and large Fermi velocity, energy level spacing in graphene nano structures  can easily be larger than the temperature, the tunneling induced broadening,  and the perturbations created by the adsorbates.  One of the consequences is that  single electron transport takes place through a single orbital level. 

 Our simulations show how the spin splitting $\delta$ of the transport level is sensitive to the average magnetization of the  magnetic adsorbates, which is controlled by application of a magnetic field along the plane of graphene, to avoid diamagnetic shifts.   On the other hand, the linear conductance  $G$ of the single electron transistor depends on $\delta$, which accounts for the sensing mechanism. More specifically, $G$  depends on $\delta$
due to either changes in the spin dependent addition energies $\Delta_{\sigma}$ or changes in the 
electrons lifetime $\Gamma_{\sigma}$. The first is independent of the nature of 
the electrodes, whereas the second only happens if they are magnetic or have an energy dependent density of states.

We have shown how, within an independent particle model and in the single electron transport regime, the energy dependence of the graphene electrode density of states can only be
relevant if  the transport energy level is fine tuned to the 
Dirac point. However, this fine tuning is quite unlike to happen in experimental conditions\cite{Geim-closest}. Still,  the combined action of disorder and Coulomb interaction could give rise to a so called Coulomb Gap in the density of states of graphene, that might  make the tunneling rates depend on the energy.\cite{Coulomb-gap-graphene,Coulomb-gap-graphene1,Coulomb-gap-graphene2} 

Finally,  we have assumed that both the edges of the graphene island and graphene electrodes are non-magnetic. Our discussion of the effect of  spin-polarized electrodes on the transport properties of the device  would be valid for electrodes with  ferromagnetic  zigzag edges.  A second possibility, out of the scope of this work, is to consider a graphene single electron transistor whose central island has ferromagnetic edges. This case has been already studied. \cite{Ezawa}

In conclusion, we have studied the mechanisms by which a graphene single electron 
transistor could work as a sensor of the magnetic order of  
 magnetic atoms or molecules adsorbed on the graphene central region.  Our work has been motivated in part by recent experimental
 works, \cite{Wernsdorfer_NL,Candini-PRB-2011}.   Whereas further work is still necessary
 to nail down the  physical mechanisms for  the spin sensing principles underlying the experimental work, our study provides a conceptual framework for graphene single electron transistor spin sensors.

%%%%%%%%%%%%%%%%%%%%% 
This work has been financially  supported by MEC-Spain (Grant Nos. FIS2010-21883-C02-01 and CONSOLIDER CSD2007-0010) as well as Generalitat Valenciana,  grant Prometeo 2012-11. We thank A. Candini for useful comments on the manuscript.

%%%%%%%%%%%%%%
%REFERENCES
%%%%%%%%%%%%%%
\bibliographystyle{apsrev}

\begin{thebibliography}{21}
  
\bibitem{Schedin07} F. Schedin, A. Geim, S. Morozov, E. Hill, P. Blake, 
M. Katsnelson, and K. Novoselov, 
Nature Materials \textbf{6}, 652 (2007).
  

\bibitem{Wernsdorfer_NL} A. Candini, S. Klyatskaya, M. Ruben, W. Wernsdorfer,
and M. Affronte, Nano Letters \textbf{11}, 2634 (2011).
   

\bibitem{Pisana} %Tunable Nanoscale Graphene Magnetometers
S. Pisana , P. M. Braganca , E. E. Marinero and B. A. Gurney, Nano Letters \textbf{10}, 341 (2010).

\bibitem{Pesin2012} D. Pesin and A. MacDonald, Nature Materials \textbf{11}, 409(2012).


\bibitem{Candini-PRB-2011}  A. Candini,  C. Alvino, W. Wernsdorfer, M. Affronte, 
% Hystersis loops of magnetocondutance in graphene devices.
Physical Review B {\bf 83}, 121401 (2011).

\bibitem{Wernsdorfer_NatMat} M. Urdampilleta, S. Klyatskaya, J. Cleuziou, M. Ruben, and W. Wernsdorfer,
Nature Materials \textbf{10}, 502 (2011).

\bibitem{Yazyev07} %Defect-induced magnetism in graphene
O. V. Yazyev and L. Helm, Physical Review B {\bf 75}, 125408 (2007).

\bibitem{Palacios2008} J. J. Palacios, J. Fern\'andez-Rossier, L. Brey, 
Physical Review B \textbf{77}, 195428 (2008).

\bibitem{Soriano2010} D. Soriano, F. Mu\~noz-Rojas, J. Fern\'andez-Rossier, J. J. Palacios, 
Physical Review B \textbf{81}, 165409  (2010).

\bibitem{Beenakker} C.W.J. Beenakker, Physical Review B \textbf{44}, 1646 (1991).

\bibitem{Geim2008} L. Ponomarenko, F. Schedin, M. Katsnelson, R. Yang, E. Hill, K. Novoselov 
and A. K. Geim, Science {\bf 320}, 356 (2008).

\bibitem{review-dot}  
%Review: Transport through graphene quantum dots 
J. G\"uttinger, F. Molitor, C. Stampfer, S. Schnez, A. Jacobsen, S. Dr\"oscher, T. Ihn and K. Ensslin,
Rep. Prog. Phys. \textbf{75}, 126502 (2012). % Reports on Progress in Physics 


\bibitem{Fede2009} F. Mu\~noz-Rojas,  J. Fern\'andez-Rossier, J. J. Palacios, 
Physical Review Letters \textbf{102}, 136810 (2009). 


\bibitem{Castro-Neto} C. H. Lewenkopf, E. R. Mucciolo, and A. H. Castro Neto, Physical Review B
\textbf{77}, 081410(R) (2008). 

\bibitem{Ezawa} % Coulomb blockade in graphene nanodisks
M. Ezawa, Physical Review B \textbf{77}, 155411 (2008).

\bibitem{JP_JR}%Magnetism in Graphene Nanoislands
J. Fern\'andez-Rossier and J. J. Palacios,
Phys. Rev. Lett. \textbf{99}, 177204 (2007).

\bibitem{Yazyev} %Emergence of magnetism in graphene materials and nanostructures 
O. V. Yazyev, Rep. Prog. Phys. \textbf{73}, 056501 (2010). 

\bibitem{Brey-Das-Sarma} 
%Diluted Graphene Antiferromagnet
L. Brey, H. A. Fertig, and S. Das Sarma,
Phys. Rev. Lett.{\bf  99}, 116802 (2007).

\bibitem{Anderson61}
%  Localized Magnetic States in Metals
P. W. Anderson Phys. Rev. {\bf 124}, 41 (1961).

\bibitem{Schrieffer-Wolff}
%Relation between the Anderson and Kondo Hamiltonians
J. R. Schrieffer and P. A. Wolff
Phys. Rev. {\bf 149}, 491 (1966).

\bibitem{Recher2000}
%Quantum Dot as Spin Filter and Spin Memory
P. Recher, E. V.  Sukhorukov, and Daniel Loss
Phys. Rev. Lett. {\bf 85}, 1962 (2000).

\bibitem{Haug_Jauho_book_1996} H. Haug and A.-P. Jauho, Quantum kinetics in transport
and optics of semi-conductors (Springer-Verlag, Berlin, 1996).

\bibitem{Ono} K. Ono, H. Shimada, and Y. Ootuka, Journal of the Physical society of Japan 
\textbf{66}, 1261 (1997).

\bibitem{Vanwees} S. J. Van Der Molen, N. Tombros, and B. J. Van Wees, Physical
Review B \textbf{73}, 220406 (2006).

\bibitem{Barnas98}
J. Barnas,  A. Fert % 1998 Magnetoresistance oscillations due to charging effects in double ferromagnetic tunnel junctions
 Phys. Rev. Lett. {\bf 80} 1058 (1998).
 
\bibitem{Seneor2007} 
P. Seneor, A.  Bernand-Mantel and F.  Petroff,
Journal of Physics: Condensed Matter {\bf 19},  165222 (2007).
%Nanospintronics: when spintronics meets single electron physics


\bibitem{Barreiro2012} 
%Quantum Dots at Room Temperature carved out from Few-Layer Graphene DOI: 10.1021/nl3036977
A. Barreiro, H.S.J. van der Zant, L.M.K. Vandersypen,  	Nano Letters  \textbf{12}, 6096 (2012).


\bibitem{Geim-closest} %Experimental paper of Geim, addressing how close can you be to the Dirac point
A. Mayorov,  D. C. Elias, I. S. Mukhin, S. V. Morozov, L. Ponomarenko, K. S. Novoselov, A. K. Geim, and R. V. Gorbachev, Nano Letters \textbf{12}, 4629 (2012).


\bibitem{Coulomb-gap-graphene} 
%Coulomb gap in graphene nanoribbons
S. Dr\"oscher, H. Knowles, Y. Meir, K. Ensslin, and T. Ihn, Physical Review B \textbf{84}, 073405 (2011). 
%

\bibitem{Coulomb-gap-graphene1} % Original Paper: Coulomb gap and low temperature conductivity of disordered systems 
%doi:10.1088/0022-3719/8/4/003
A. L. Efros, and B. I. Shklovskii, J. Phys. C: Solid State Phys. \textbf{8} 49 (1971). 
%

%
\bibitem{Coulomb-gap-graphene2} % Disorder induced Coulomb gaps in graphene constrictions with different aspect ratios 
B. Terr\'es, J. Dauber, C. Volk, S. Trellenkamp, U. Wichmann, and C. Stampfer, 
Applied Physics Letters \textbf{98}, 032109 (2011). 

\bibitem{dots_Ensslin}
% Electrostatic confinement of electrons in graphene nanoribbons
X. Liu, J. B. Oostinga, A. F. Morpurgo, and L. M. K. Vandersypen, 
Physical Review B \textbf{80}, 121407(R) (2009).  


\bibitem{Sols-Guinea} %Coulomb Blockade in Graphene Nanoribbons
F. Sols, F. Guinea, and A. H. Castro Neto, 
Physical Review Letters \textbf{99}, 166803 (2007).


\end{thebibliography}

\end{document}